\documentclass[aps,pre,twocolumn]{revtex4}

\usepackage[english]{babel}
\usepackage[utf8]{inputenc}
\usepackage{graphicx}
\usepackage{mathtools}
\usepackage{amsfonts}
\usepackage{amsmath}
\usepackage{url}	
\usepackage{subcaption}
\usepackage{booktabs}
\usepackage{textcomp}
\usepackage{listings}
\usepackage{xcolor}
\usepackage{mhchem}
\usepackage{wrapfig}
\usepackage{adjustbox}
\usepackage{siunitx}
\usepackage{empheq}
\usepackage{tablefootnote}
\usepackage{fancyhdr}
\usepackage{titlesec}
\usepackage{siunitx}
\usepackage{svg}
\usepackage[pscoord]{eso-pic}
\usepackage{booktabs}
\usepackage{array}
\usepackage{multirow}
\usepackage{graphicx}
\usepackage{comment}

\makeatletter
\newcommand{\thickhline}{%
    \noalign {\ifnum 0=`}\fi \hrule height 1pt
    \futurelet \reserved@a \@xhline
}
\newcolumntype{"}{@{\hskip\tabcolsep\vrule width 1pt\hskip\tabcolsep}}
\makeatother

\titleformat{\section}{\normalfont\scshape}{\arabic{section}}{1em}{}
\titleformat{\subsection}{\normalfont\scshape}{\arabic{section}.\arabic{subsection}}{1em}{}
\titleformat{\subsubsection}{\normalfont\scshape}{\arabic{section}.\arabic{subsection}.\arabic{subsubsection}}{1em}{}

\begin{document}

\title{Thermal effects and spontaneous frictional relaxation in atomically thin layered materials}
\author{J Roadnight Sheehan$^{2}$}
\author{David Andersson$^{1,2}$}
\author{Astrid S.\ de Wijn$^{2,1}$}
\email{astrid.dewijn@ntnu.no}
\affiliation{$^1$ Chemical Physics Division, Department of Physics, Stockholm University, Sweden}
\affiliation{$^2$ Department of Mechanical and Industrial Engineering, Norwegian University of Science and Technology, Norway}

\begin{abstract}
We study the thermal effects on the frictional properties of atomically thin sheets.  We simulate a simple model based on the Prandtl-Tomlinson model that reproduces the layer dependence of friction and strengthening effects seen in AFM experiments.  We investigate sliding at constant speed as well as reversing direction.
We also investigate contact ageing: the changes that occur to the contact when the sliding stops completely.  We compare the numerical results to analytical calculations based on Kramers rates.  We find that there is a slower than exponential contact ageing that weakens the contact and that we expect will be observable in experiments.
We discuss the implications for sliding as well as ageing experiments.
\end{abstract}

\maketitle

\section{Introduction}
During the past decade, atomically thin sheets consisting of single or multiple layers of a 2D material have been the subject of much investigation \cite{sheets1,sheets2} and are an important component in nanotechnology in general \cite{grapheneapps,2dmats}; as well as a promising candidate as friction reducing agents in microscopic as well as macroscopic applications \cite{graphlube}. The frictional properties of atomically thin sheets have been found to exhibit several interesting phenomena.  Specifically, friction decreasing with increasing number of layers, and an initial friction strengthening. These were first observed in AFM experiments over a decade ago \cite{carpick1,filleter}.  To further the understanding of these materials, atomistic MD simulations have also been performed by a number of groups, (such as \cite{carpick2,martinipaper}).  However, significant developments were recently made in this area using a simple model~\cite{paper1}.

While much understanding has been gained of the strengthening and layer-number-dependence of friction of layered-materials, one aspect that is yet to be understood or investigated is how friction, in such systems, is impacted by thermal effects. Thermal fluctuations play an important role in experimental nano-scale friction, especially for systems like these, where the energy barriers against sliding tend to be relatively low~\cite{sang,thermolubricity}. Conversely, thermal effects in general in layered materials have been the subject of much scrutiny (see, for example~\cite{annalise,pop2013thermal}).

In this work, we analyse the thermal behaviour of the phenomena of strengthening and layer dependence of the friction of atomically thin sheets using the simple modified Prandtl-Tomlinson (PT) model\cite{pt1,pt2,pt3} that has previously been shown to capture and explain these phenomena~\cite{paper1}.  We add thermal noise to this model, and investigate its effect on the distortion of the sheets and the friction. Furthermore we investigate how contact ageing results from thermally activated relaxations in the system. Contact ageing -- when the strength of the contact changes during sticking periods -- is an established phenomenon that occurs on wide range of length scales from the atomic (see, for example~\cite{aging1,aging2,aging3}) to the geological ~\cite{aging4,aging5}.  Ageing usually leads to increasingly strong contacts, and thus plays a role in the phenomenological (macroscopic) observation that static friction is typically larger than dynamic friction. Simple extensions to the PT-model has been used before to investigate the effect of aging on friction, see e.g. \cite{refwantedmetocitethispaper}, the novelty in the present work however is that the model geared towards atomically thin layered materials. 

We will begin by introducing and quickly reviewing the model proposed in~\cite{paper1}. Following this, we will discuss how the thermal fluctuations affect the dynamics during sliding. The main body of this report will be devoted to studying thermal relaxations after sliding has stopped: an AFM tip that has been scanning and deforming a sheet for some distance is stopped; then, due to thermally activated decay, the system slowly re-equilibrates as the sheet relaxes. We combine this with analytical calculations of thermal activation using Kramers theory, which we show can calculate the most probable decay paths in the complex potential landscape, and thus predict estimated relaxation times.

An interesting consequence of the structure of the potential landscape is that relaxation times appear over many orders of magnitude, which implies that it should be possible to observe these effects on experimental time scales.  It also shows that the contact ageing is slower than exponential, which is typically what is observed in experiments~(see e.g.~\cite{aging1}).  What is however more unusual is that the ageing actually weakens the contact.

\section{The model and simulation setup\label{sec:methods}}

\begin{figure}
    \centering
    \includegraphics[width=1\linewidth]{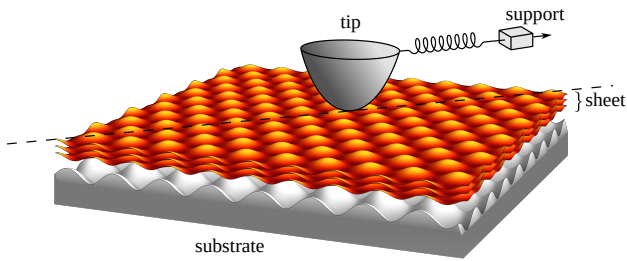}
    \caption{Illustration of the system and model setup~\cite{paper1}. The model is similar to the PT model, with a support pulling the tip.  A single additional variable is introduced to describe the state of distortion of the entire sheet. Figure from \cite{paper1}, shared under the Creative Commons Attribution 4.0 International License \cite{licence}.}
    \label{fig:SimpleModel}
\end{figure}

\subsection{A simple model for the friction of atomically thin sheets}

The model we consider here was first presented in Ref.~\cite{paper1}, and is an extension of the traditional one-dimensional PT model.  It is based around the setup shown in Fig.~\ref{fig:SimpleModel}.  

As in the PT model, the tip (position $x$) is pulled via a spring by a support moving at a constant velocity, $v$.  The tip is subject to viscous damping and the system has a potential energy that describes the interaction between the tip and the material it slides on.  To describe the atomically thin sheets between the tip and the substrate, this extended model introduces an extra degree of freedom, $q$, which describes the distortion in the sheet. Here, distortion means any change in the sheet or sheets; this can be bending, shearing, shifting, or very local displacements of individual atoms. The total potential energy of the system is then described as a function of both the tip and sheet via
\begin{equation}
\begin{split}
\label{eqn:Potential}
U(x,q,t)=\frac{k}{2}(x-vt)^2+V_\mathrm{sheet}(q)+\\
V_\mathrm{tip-sheet}(x,q)+V_\mathrm{tip-substrate}(x,q)~.
\end{split}
\end{equation}
The first term in the potential describes the stiffness of the tip and AFM cantilever with a single spring constant $k$.
The other terms in the total potential energy are: the internal potential energy of the sheet $V_\mathrm{sheet}$; the interaction between the tip and the sheets $V_\mathrm{tip-sheet}$; and the interaction between the tip and the substrate through the sheet $V_\mathrm{tip-substrate}$.

The terms in the potential energy related to the sheet and substrate can be written in a general way by including leading order, and in some cases next-to-leading order, terms in an expansion in $q$~\cite{paper1}. Higher orders were found to give no extra insight or relevance to the problem. This yields
\begin{eqnarray}
\label{eqn:Vsheet}
V_\mathrm{sheet}(q)=\nu_2q^2+\nu_4q^4~,\\
\label{eqn:Vtipsheet}
V_\mathrm{tip-sheet}(x,q)=\big(V_1+\kappa_1q^2\big)\bigg(1-\cos\bigg[\frac{2\pi}{a}(x-q)\bigg]\bigg)~,\\
\label{eqn:Vtipsubstrate}
V_\mathrm{tip-substrate}(x,q)=\big(V_2+\kappa_2q^2\big)\bigg(1-\cos\bigg[\frac{2\pi}{b}x\bigg]\bigg)~,
\end{eqnarray}
where $\nu_2$ and $\nu_4$ are parameters describing the potential energy from distortion of the sheet; $V_1$ and $V_2$ are the potential-energy corrugation of the tip on an undistorted sheet and the contribution of the substrate respectively; $\kappa_1$ and $\kappa_2$ account for changes in the corrugation; and $a$ and $b$ are the lattice parameters of the sheets and substrate respectively.

The extra degree of freedom for distortion, $q$, has both a global and local effect on the potential landscape. The local effect is created by the local interaction with the tip through the cosine term in equation \ref{eqn:Vtipsheet}; whereas all other instances of $q$ give rise to global effects in the potential landscape of the system.

This model captures a number of typical frictional behaviours exhibited by layered materials in AFM experiments.  Specifically, it displays the strengthening that appears as an initial increase in the force until the regular steady-state stick-slip pattern begins, as well as the dependence of the friction on the thickness of the sheet. These characteristics of the friction of layered materials have been reproduced many times in both AFM experiments and molecular dynamic simulations~\cite{carpick1,carpick2}.

A crucial role in the dynamics of this model is played by the position of the energy minima in the potential-energy landscape.  In the low-velocity limit, the system is quasi-static, and the tip is pulled from minimum to minimum, initially distorting the sheet and then later sliding along it.  For more details, see Ref.~\cite{paper1}.

\begin{figure}
    \includegraphics[width=8.6cm]{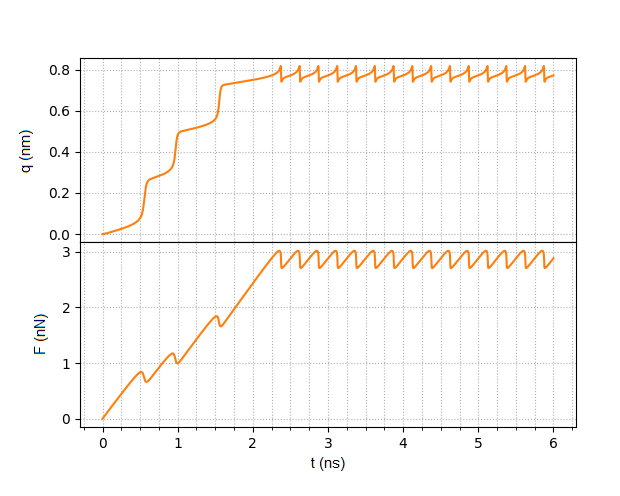}
    \\[-5.8cm](a)\hfill\strut\\[5.3cm]
    \includegraphics[width=8.6cm]{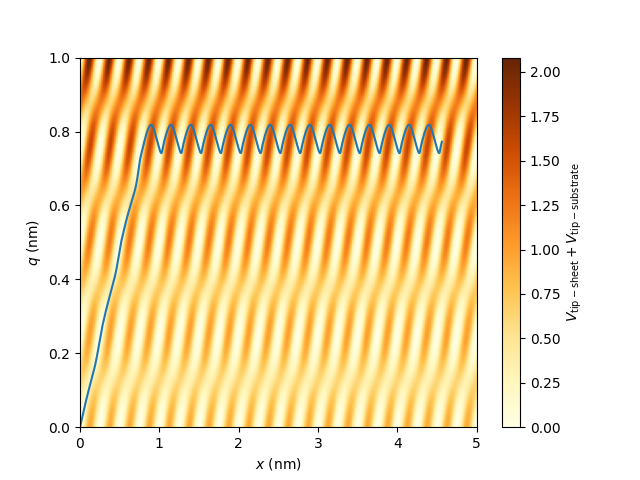}
    \\[-5.5cm](b)\hfill\strut\\[5.0cm]
    \caption{The behaviour of the system without temperature: (a) the force and distortion, $q$, as a function of time and (b) the trajectory in the $xq$ plane, superimposed on the periodic terms in the potential-energy landscape.  The rest of the parameter values are given in the text. The behaviour of the system is controlled by the minima in the potential-energy landscape.
    Figure from \cite{paper1}, shared under the Creative Commons Attribution 4.0 International License \cite{licence}.}
    \label{fig:nonthermal}
\end{figure}

In Fig.~\ref{fig:nonthermal} we show a baseline summary of how the model behaves in the absence of thermal noise. It shows the strengthening, and in the $xq$ plot one can see how this is related to the structure of the energy landscape.
After the strengthening, during steady-state sliding, the tip remains in a small range of $q$ and stays near one row of minima in the potential-energy landscape. This is explored in some detail in \cite{paper1}, and is reflected in Fig.~\ref{fig:nonthermal}. This results in a periodic steady-state, that looks very similar to the one-dimensional PT model.

In this work, we introduce thermal fluctuations into this model using Langevin dynamics, i.e.\ in addition to the viscous damping with damping constants $\eta_x$ and $\eta_q$ in $x$ and $q$ respectively, we include Gaussian random forces $\xi_x(t)$ and $\xi_q(t)$.
The equations of motion then read
\begin{eqnarray}
        \label{eq:eom1}
        -m_x \ddot{x} = - \frac{\partial U(x,q)}{\partial x} - m_x \eta_x \dot{x} + A_x \xi_x(t)~, \\
        \label{eq:eom2}
        -m_q \ddot{q} = - \frac{\partial U(x,q)}{\partial q} - m_q \eta_q \dot{q} + A_q \xi_q(t)~,
\end{eqnarray}
where $m_x$ and $m_q$ are the effective masses of the tip and sheet distortion respectively.  The thermal noise is given by the amplitudes $A_{x,q} = \sqrt{2k_B T \eta_{x,q}}$ and Gaussian uncorrelated noises $\xi_x(t)$ and $\xi_q(t)$  with standard deviation $1$ and 0 mean.

Unless otherwise mentioned, the parameter values we use are the same as in our previous work~\cite{paper1}.  They were estimated from detailed molecular-dynamics simulations that showed strengthening and layer-dependence~\cite{carpick2} and that modelled an AFM tip sliding over layers of graphene.  In some cases of missing information, we have used AFM experiments as a starting point and picked our values based on this. The parameter value are as follows. The masses are $ m_x=10^{-23} \:\si{kg} $ and $ m_q= 3.67 \times 10^{-24} \:\si{kg} $, which are in the order of magnitude thousands of carbon atoms, similar to that used by previous authors in this context \cite{carpick2}. The damping coefficients are $ \eta_x = 18.75 \:\si{ps^{-1}} $ and $ \eta_q = 42.86 \:\si{ps}^{-1}$, which are typical values for an AFM. For the tip-support interaction: $v = 1.0 \:\si{ms^{-1}}$, $k = 2.0 \:\si{Nm^{-1}} = 12.5\:\si{eVnm^{-2}}$ and $a = 2.5 \:\si{\angstrom}$; the support velocity is similar to that used in MD simulations, the spring constant is on the order of magnitude of an ordinary cantilever, and the lattice spacing is that of graphene when scanning in the geometrically regular, zig-zag, direction. To keep matters simple, we use a commensurate combination of lattice parameters, $ a/b=1$. The corrugation parameters are $V_1 = 0.31 \:\si{eV}$ and $V_2 = 0.15 \:\si{eV}$, which are common energy levels use d in the ordinary PT model. The corrugation coefficients: $ \kappa_1 = 0.375 \:\si{eVnm^{-2}}$ and $\kappa_2 = 0.1875 \:\si{eVnm^{-2}}$, were picked on the same grounds as the corrugation parameters. The distortion energies are chosen to be $\nu_2 = 2.39 \:\si{eVnm^{-2}}$ and $\nu_4 = 3.64 \:\si{eVnm^{-4}}$, these represent binding energies between the sheet and the substrate as well as the internal bonding energies in the sheet respectively, and their values were picked accordingly. 

The exact parameter values are not critical for the model to qualitatively reproduce friction dynamics of layered materials. Rather the model works for a wide range of parameter values. We will show however how the $V$ and $\kappa$ energy parameters change the barrier heights of the potential energy landscape, and how $\nu$ parameters restricts the accessibility of the corresponding potential minima. Crucially, $\nu_4$ needs to be large enough to stabilize the $q$ variable, while still allowing some freedom for $q$ to change.

\subsection{Numerical simulations}

The equations of motion [Eqs.~(\ref{eq:eom1}) and~(\ref{eq:eom2})] were integrated using a fourth order Runge-Kutta algorithm implemented in \verb|C++| with timestep $15\:\si{fs}$. This timestep size was verified by checking energy conservation, as well as giving a reliable sampling frequency even during slips. 

For sliding simulations we choose as initial conditions an undistorted sheet ($q=0$) and the tip and support both at $x=vt=0$, with tip velocity $\dot{x}=0$.
Along with studying the trajectories, we investigate the thermal relaxation after sliding has stopped. In this case, as $v=0$, we refer to the support position as $r$ in this stationary case. For this purpose, we collect statistics from a large number of simulations.  To simplify the analysis, we start each of these simulations from the same initial conditions.
We use as initial conditions the zero-temperature slipping point in the steady state sliding regime. This point was used as initial state for the relaxation simulations. This corresponds to the following values: $x=1.03\:\si{nm}$, $\dot{x}=1.00\:\si{ms^{-1}}$, $\ddot{x}=1.86\times10^{13}\:\si{ms^{-2}}$ , $q=0.74\:\si{nm}$, $\dot{q}=0.51\:\si{ms^{-1}}$, $\ddot{q}=2.17\times10^{13}\:\si{ms^{-2}}$ and $r=2.448\:\si{nm}$.
While this is not a procedure that can be repeated in an experiment, as we will discuss later on, the aspects of the decay that would be experimentally detectable are not affected.

We identify decay points using a simple procedure. To eliminate the risk of large, short-range thermal fluctuations being picked up as decays, we first reduce the noise by smoothing it with a running mean over 1000 timesteps ($15\si{ns}$). From the potential energy we can calculate all the minima in the potential landscape. Furthermore, we know by construction that the tip starts in the minimum labeled $m1$. We can define a proximity measure: 
\begin{equation}
    d = \sqrt{ (x-x_0)^2 + (q-q_0)^2},
\end{equation}
 of the tip to each minimum with some coordinates $(x_0,q_0)$. The algorithm at every timestep checks if the tip is sufficiently close to any minimum ($d < 0.025$ nm). If that is the case, then it checks if this minimum is the same minimum that it was assigned to previously. If that is not the case it further checks if the tip has been closer to this minimum than any other for ten consecutive timesteps. If that is the case, the algorithm determines that a decay has occurred into the new minimum at the time all these conditions were first met.

\section{Simulation results}\label{SimRes}

The thermal fluctuations in this distortion can influence the friction and other behaviour of the system in several ways.  In this section, we first investigate numerically the effects during normal sliding, and then move onto the phenomena that appear when the sliding is reversed or stopped.

\subsection{Sliding}

\begin{figure}
    \includegraphics[width=8.6cm]{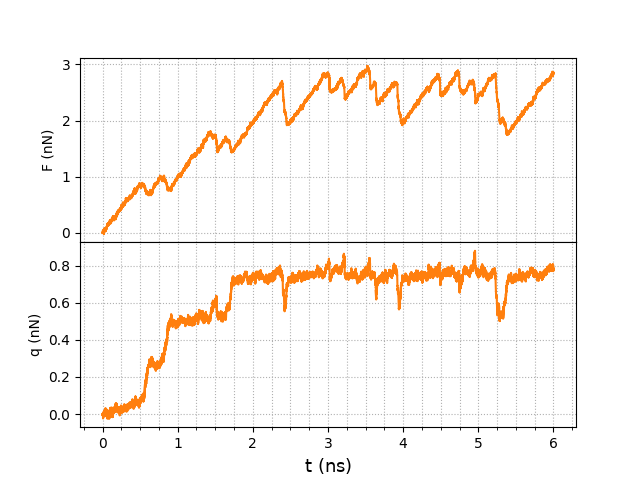}
    \\[-5.5cm](a)\hfill\strut\\[5.0cm]
    \includegraphics[width=8.6cm]{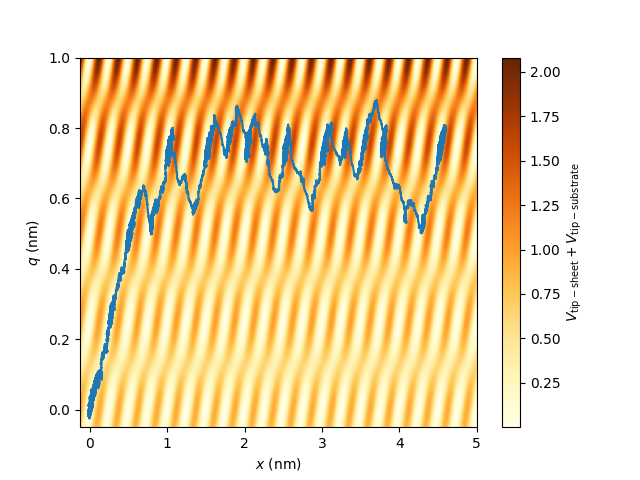}
    \\[-5.5cm](b)\hfill\strut\\[5.0cm]
    \caption{The behaviour of the system with temperature: (a) the force and distortion, $q$, as a function of time and (b) the trajectory in the $xq$ plane, superimposed on the periodic terms in the potential-energy landscape.  The rest of the parameter values are given in the text. We see an irregular stick-slip pattern due to thermal noise, and the system sometimes randomly changes to a lower value of $q$.}
    \label{fig:forcethermal}
\end{figure}

In Fig.~\ref{fig:forcethermal}, we show a typical example of the evolution of a force and $q$ trace, and a path taken in the underlying potential-energy landscape by the tip.  We observe that besides the normal irregularity in the force trace, which one would expect due to thermal noise, there is an additional pattern of larger short-time decreases in the force corresponding to temporarily lower values in $q$: when the tip falls into a minimum on a row closer to $q=0$.  This corresponds to lower corrugation of the sheet and thus lower lateral force.

The model displays a thermal behaviour that is similar to the one-dimensional PT model \cite{ptthermal}.
The random thermal noise kicks the tip over the barrier into the next minimum of the substrate earlier than it would have moved without the noise. This leads to slips at a lower lateral force, and thus lowers friction. This phenomenon is called thermolubricity, see \cite{sang} for a formal treatment, and ~\cite{thermolubricity} for experimental support within the context of a one-dimensional PT system. Since the present model has one extra degree of freedom compared to the PT model, the thermal behaviour is more complex.
There are, as in the one-dimensional PT model, thermally activated slips in the $x$-direction. However, the tip can also overcome the barrier separating two rows of minima, and briefly switch to a row of minima with a lower distortion, $q$.  During the next slip, the strengthening process once again pulls it up to its native row. This is responsible for a very typical dip in the lateral force and is a signature of the presence of the sheet that might be observable in experiments.

\begin{figure}
        \includegraphics[width=8.6cm]{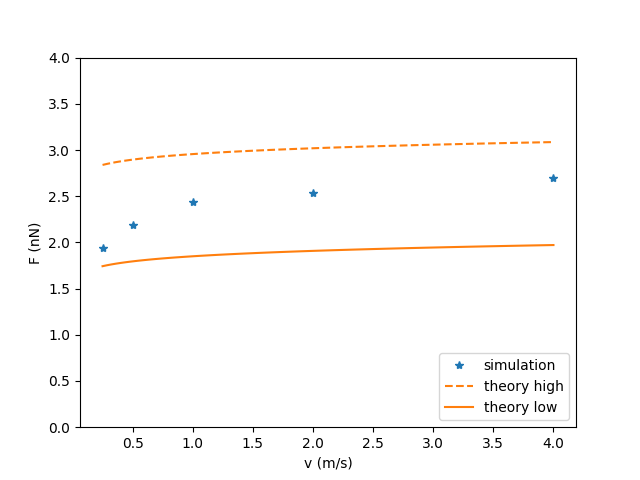}
        \\[-5.5cm](a)\hfill\strut\\[5.0cm]
        \includegraphics[width=8.6cm]{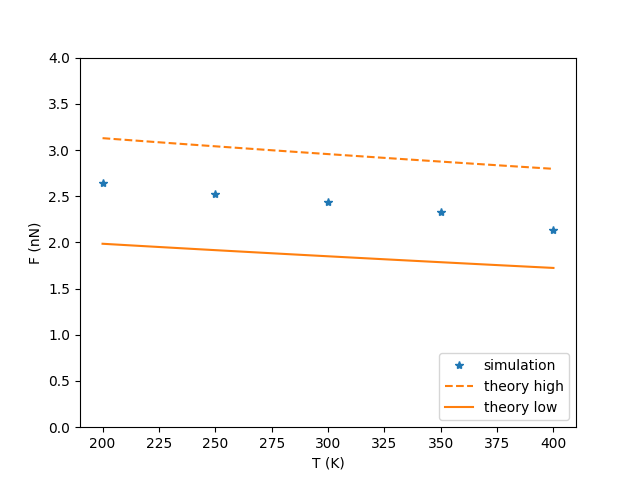}
        \\[-5.5cm](b)\hfill\strut\\[5.0cm]
    \caption{The friction as a function of (a) velocity and (b) temperature, along with theoretical estimates based on a constant value of $q$ and Sang's Kramers rate treatment~\cite{sang} of the one-dimensional PT model, given in Eqn.~(\ref{eqn:sangforce}).}
    \label{fig:sangcomp}
\end{figure}

In the one-dimensional PT model with nonzero temperature, thermal activation of slips can be handled using Kramers rate theory, resulting in a friction that depends on the corrugation of the substrate, the velocity, and temperature through~\cite{sang}
\begin{equation}
    F=F_\mathrm{c}-\Delta F |\ln v^*|^{2/3},
    \label{eqn:sangforce}
\end{equation}
where $F_\mathrm{c}$ is the lateral force in the inflection point of the substrate potential in the absence of thermal noise, $v^*$ is a dimensionless velocity that depends on the support velocity as well as temperature, and $\Delta F$ is a constant. Both $F_\mathrm{c}$ and $\Delta F$ involve the corrugation of the substrate.  For a formal treatment, as well as full expressions for these constants we refer to~\cite{sang}. 

We can estimate the friction based on these one-dimensional expressions by reducing our system to one-dimension. This is done by approximating $q$ as a constant and obtaining an approximate corrugation from that.  Previously, we showed that $q$ in the steady state could be estimated reasonably well from its maximum attainable value, $q_\mathrm{max}$, given by the sole real root to the polynomial: $2\nu_2 q_\mathrm{max}+4 \nu_4 q_\mathrm{max}^3 + 2 \kappa_1 q_\mathrm{max} - \frac{2 \pi}{a} (V_1 + \kappa_1 q_\mathrm{max}^2) = 0$~\cite{paper1}.
From Eqn.~(\ref{eqn:Potential}), we can see that the corrugation potential then has two terms, one from the sheet and one from the substrate, with corrugations $U_1=V_1 + \kappa_1 q_\mathrm{max}^2$ and $U_2=V_2 + \kappa_2 q_\mathrm{max}^2$ respectively.
The tip slips between minima of both terms, and must cross barriers inbetween, which lie on the ridge of maximum tip-sheet energy.  We can estimate the effective corrugation by considering the energy difference between the minima (where both terms are at a minimum) and the points on the ridges where the tip-sheet term has a maximum, but the tip-substate term can have any value between $-U_2$ and $+U_2$.  This gives lower and upper bounds for the effective corrugation of
\begin{eqnarray}
    V_\mathrm{lower} = V_1 + \kappa_1 q_\mathrm{max}^2~,\\
    V_\mathrm{upper} = V_1 + \kappa_1 q_\mathrm{max}^2 + V_2 + \kappa_2 q_\mathrm{max}^2~.
\end{eqnarray}
Figure~\ref{fig:sangcomp} shows the friction as a function of velocity and temperature computed from the simulations, as well as the estimated lower and upper bounds for the friction.  In the simulations, the friction was obtained as the average lateral force from 120 ns simulations, excluding the strengthening period. The observed friction is within the estimated limits, but the range between our upper and lower bound is quite wide, and the quantitative behaviour is also somewhat different.  This is not surprising given that there is a plethora of new dynamics in the $q$ degree of freedom that is unaccounted for, including, for example, the hopping between different rows of minima that can be seen in Fig.~\ref{fig:forcethermal}.

\subsection{Reversing direction}

\begin{figure}
        \includegraphics[width=8.6cm]{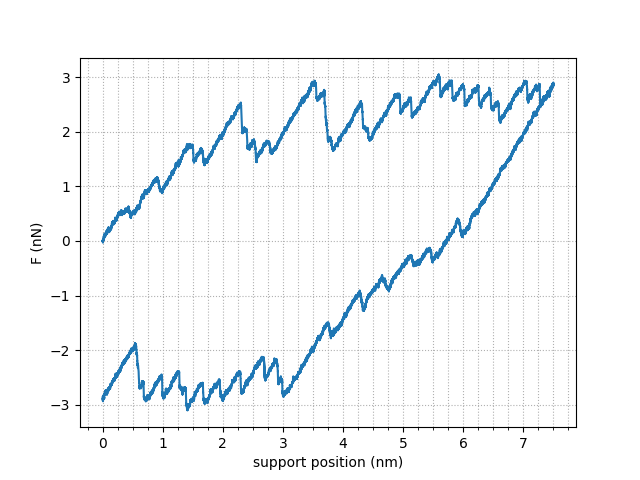}
            \\[-5.5cm](a)\hfill\strut\\[5.0cm]
        \includegraphics[width=8.6cm]{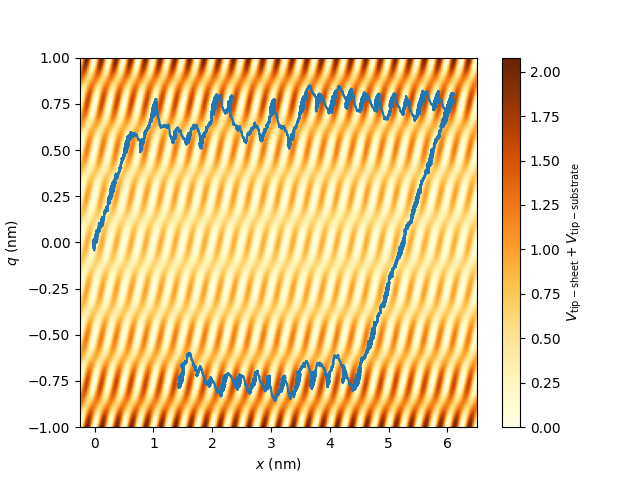}
            \\[-5.5cm](b)\hfill\strut\\[5.0cm]

    \caption{Retrace in the force (a) as well as in the $x$-$q$ parametric curve (b). Retrace curves are less likely to come back to their exact starting location in the two dimensional PT model, because the potential landscape allows for more complex thermal effects.}
    \label{fig:retrace}
\end{figure}

We now briefly consider the transient effects when the sliding direction changes.
This is common in experiments to measure friction loops, where the direction of sliding is reversed halfway through. 
A simulation of such a force trace is shown in Fig.~\ref{fig:retrace}. The parameter values in this figure are as with sliding.  The sliding direction is reversed after $2.5\:\si{ns}$, after which time the support retraces the same path it came. The distortion of the sheet $q$ does not immediately disappear when the tip changes sliding direction, and so signatures of the distortion generated by the first half of the loop are visible during the beginning of the second half in the form of a double strengthening regime.
This friction evolution is in good agreement with that found in detailed MD simulations and experiments on layered materials~\cite{carpick2,carpick1,filleter}.

\subsection{Relaxation after stopping}

\begin{figure}
    \centering
    \includegraphics[width=8.6cm]{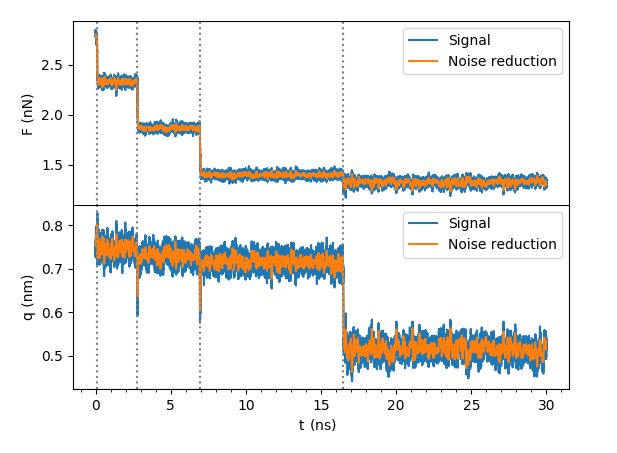}
            \\[-5.5cm](a)\hfill\strut\\[5.0cm]
    \includegraphics[width=8.6cm]{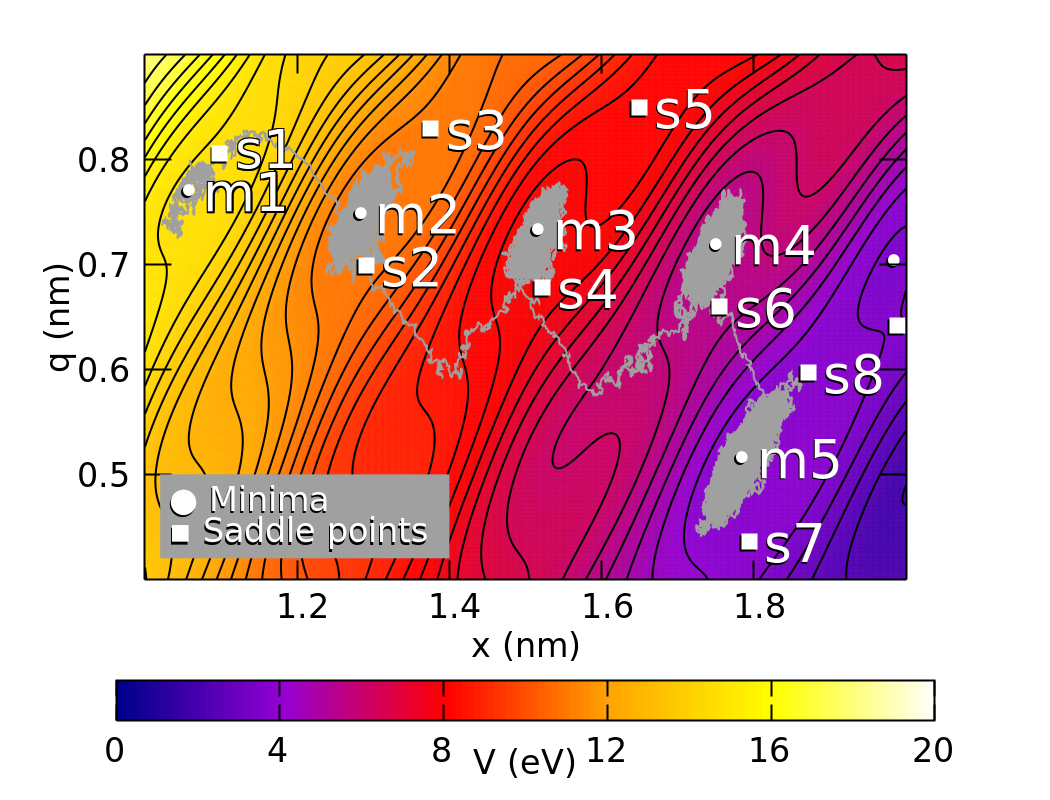}
            \\[-5.5cm](b)\hfill\strut\\[5.0cm]
    \caption{(a) The evolution in the force trace and in $q$ of the tip on a relaxing sheet. The dashed lines indicate decays. Noise filtering by running average was applied. Relaxation measured for 30 ns at 300 K. (b) The corresponding path (grey line) of the same data in relation to minima in the potential energy landscape. Contours are spaced every 0.4 eV to give a better guide of the shape of the landscape.}
    \label{fig:relaxfig}
\end{figure}

We now turn to the ageing effects that appear after the sliding stops.
An example of this is shown in Fig.~\ref{fig:relaxfig}.  Once the tip has stopped, the force decays by finite increments, as the distortion also decreases and the sheet relaxes.  This effect can again be understood by considering the potential-energy landscape.
At this point, the distortion variable, $q$, is non-vanishing, and thus the system is not in a global energy minimum, but only in a local minimum. 
The thermal noise provides random kicks to the tip, allowing it to overcome potential energy barriers by thermal activation, and move into a minimum with a lower potential energy.  This decay is a type of contact ageing.

Since we are interested in understanding what signatures of this might be visible in experiments through changes in the lateral force  we investigate the characteristics of this process of relaxation in more detail in terms of the path that the tip is likely to take across the landscape and the lifetime of each decay out of each minimum on its journey.

In the standard one-dimensional PT model, the system would relax down to a potential minimum dependent on the position of the support. However, as can be seen from Fig.~\ref{fig:relaxfig}, the added degree of freedom, $q$, also relaxes. In general, the global minimum does not correspond to a completely relaxed sheet ($q=0$), but the sheet remains somewhat distorted. Only for very specific support positions does the global minimum correspond to a fully relaxed sheet.  We have chosen a position for stopping the support that is typical, and does not produce any special geometry in the potential-energy landscape.

The precise path that the tip takes through the minima depends on the locations of the minima and barriers between them.  While the spring force makes obvious slips towards the lowest minima, the initial values of $q$ remain largely unchanged, as can be seen in Fig.~\ref{fig:relaxfig}.
This will be further examined below in subsection~\ref{geometry}.
Another immediately notable feature of Fig.~\ref{fig:relaxfig} is that a slip in the force is often accompanied by a quickly reversed slip in $q$. This will be explained in further detail later in section~\ref{kramers}.

\begin{figure}
    \centering
            \includegraphics[width=8.6cm]{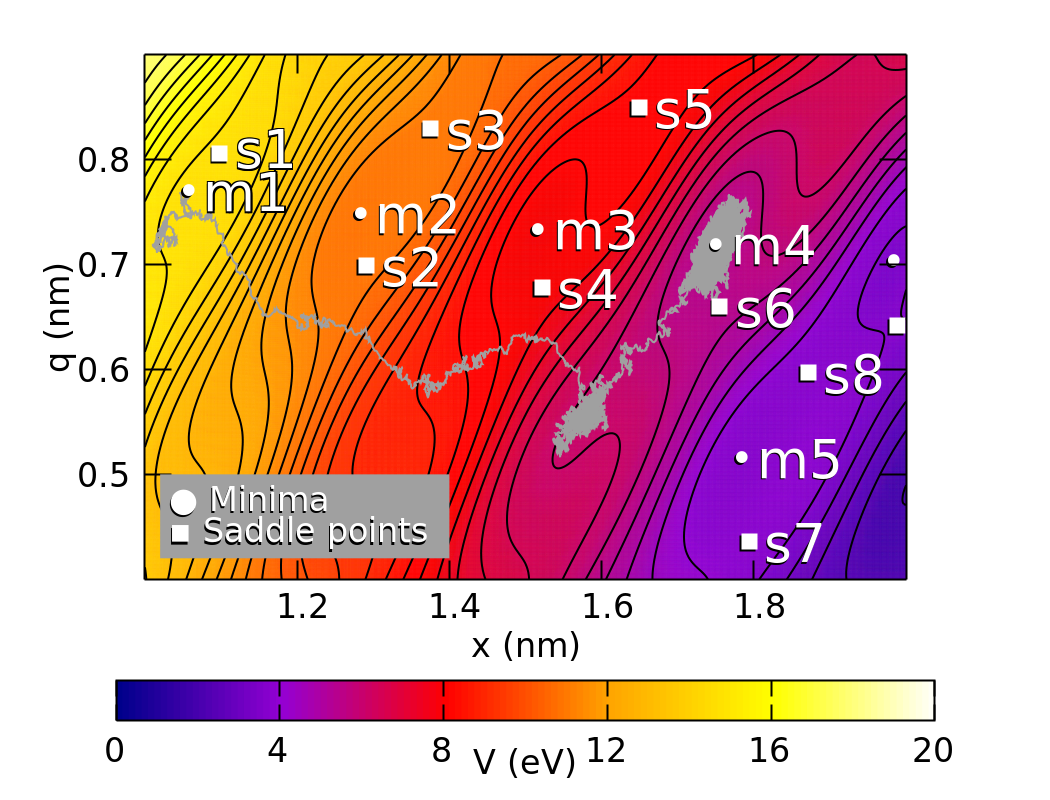}
            \\[-5.5cm](a)\hfill\strut\\[5.0cm]
            \includegraphics[width=8.6cm]{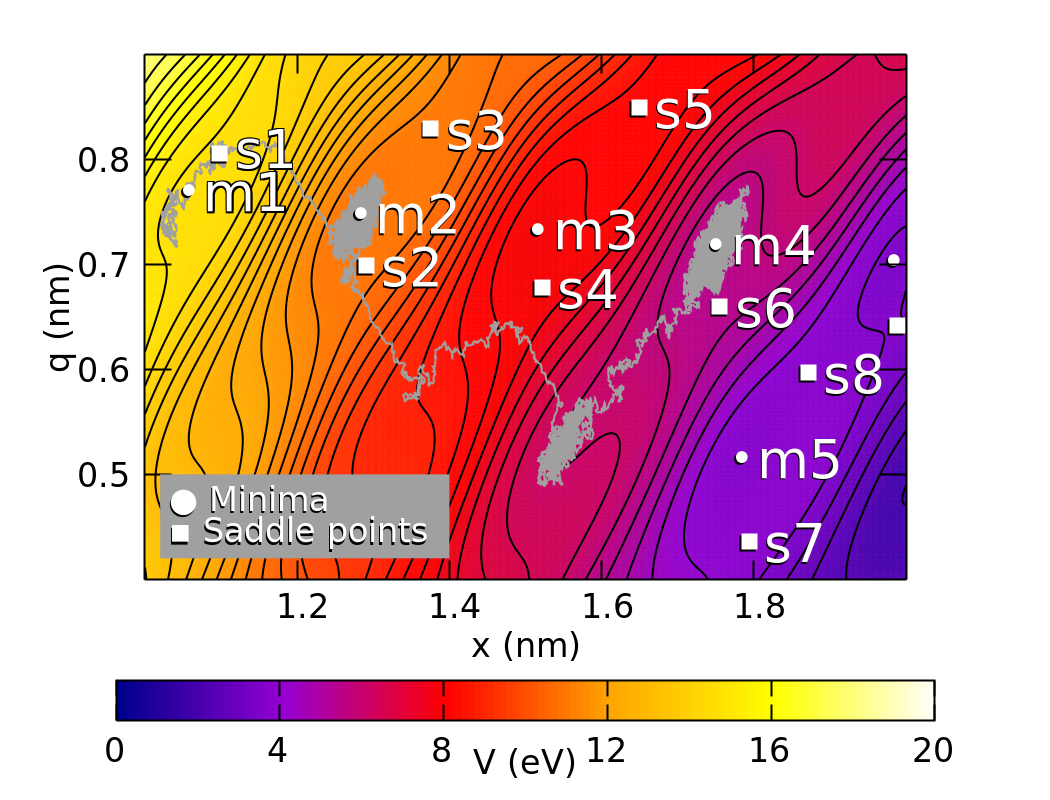}
            \\[-5.5cm](b)\hfill\strut\\[5.0cm]
            \caption{Two further example trajectories (grey lines) of the tip on the energy landscape during the first 30~ns of relaxation at 300~K. Contours are spaced every 0.4 eV to give a better guide of the shape of the landscape.}
    \label{fig:DecayPath}
\end{figure}

100 30~ns simulations of relaxation with the same initial conditions were also generated and examined. Two of the tip trajectories from this data set can be seen in Fig.~\ref{fig:DecayPath} and will be discussed in section~\ref{StrangePaths}.

\section{Theoretical considerations and calculations of the decay rates}

We now look more closely at the decay and discuss analytical approaches for better understanding the ageing.
The computational simulations of the model can show us what happens at the beginning of the decay, but due to limitations in computation time, we cannot determine numerically what will happen to the system after more than around a $\mu$s of waiting, and we cannot explore the entire space of system-specific parameters.  We can however obtain insight from the analytical calculations discussed in this section.

\subsection{Geometry of the potential-energy landscape\label{geometry}}

The ageing of the system is controlled by thermally activated decay and by the available metastable minima in the potential-energy landscape. 
We must therefore consider the general features of the geometry of these minima.  
An example of the total potential-energy landscape for the parameter values we have used in the simulations is shown in Fig.~\ref{fig:PotMinSad}, with black contour lines signifying level curves.  To understand its structure, we must look at the various contributions to the potential energy separately.

As was discussed in the previous work~\cite{paper1} and can be seen, for example, in Fig.~\ref{fig:nonthermal}, the minima of the tip-sheet-substrate system sit more or less in a grid defined by the cosine terms in Eqns. \ref{eqn:Vtipsheet} and \ref{eqn:Vtipsubstrate} for $V_\mathrm{tip-sheet}$ and $V_\mathrm{tip-substrate}$.  However the $V_\mathrm{sheet}$ term of the potential energy (Eqn.~\ref{eqn:Potential}) parabolicly warps the potential landscape around $q=0$, while the support term in a similar fashion warps the landscape around the support position (indicated by $r$). This introduces a global 2d potential well centred on $x=r, q=0$.  This removes any minima that are too far away from this point and shifts the rest towards it.
Due to the higher corrugation for large absolute values of $q$, more minima survive further away from $q=0$.

For each minimum there are at most eight nearby minima. Double decays may be possible.  However, overcoming of the initial potential barrier to escape the minimum is the same regardless of how many minima or saddle points the tip goes through during the rest of the decay.

The position of the stopped support in the lattice unit cell impacts the position of the global minimum and structure of the other local minima around it, and thus the behaviour of the system in the final stages of relaxation.
The support may be positioned such that there is a simultaneous minimum for the tip-support interaction and the tip-sheet interaction for an undistorted sheet. 
This is the case when the support position is almost equal to an integer, $n$, multiple of the lattice constant, $r\approx na$.  In this situation, the system has a possibility of nearly relaxing fully to where the tip position aligns with the support's position and the sheets have relaxed to their lowest energy so that the final state is close to $x=r,q=0$.
Conversely, the support may have stopped close to a maximum in the tip-sheet interaction for an undistorted sheet.  In this case, the global minima to which the system may relax will in general be further away from $x=r$ or $q=0$, and the tip and sheet will not be able to relax completely simultaneously.
The most extreme case of this is the most uncongruous support position, $r\approx (n+\frac12) a$.

\subsection{Calculations of the relaxation rates using Kramers theory\label{kramers}}

The relaxation path once the tip has stopped is long and consists of many steps.  Since the simulations can only show us what happens in the beginning of the decay, in order to understand better the relaxation and the path that the system takes, we calculate the relaxation rates for each possible step along the path.  We investigate the escape time of the tip from each metastable minimum through nearby saddle points during relaxation.

The decay rate for each step is determined by the local structure of the potential-energy landscape and the temperature.  Here, we use these to calculate the escape rate from one minimum through a saddle point using Kramers' theory~\cite{kramers,langer,kramersreview}.

The Kramers escape rate is given by:
\begin{equation}
\label{eqn:kramerrate}
\frac{1}{\tau}=\frac{\lambda_+}{2\pi} \sqrt\frac{\det \mathcal{E}_\mathrm{m}}{|\det \mathcal{E}_\mathrm{s}|} \exp\bigg(-\frac{\Delta U}{k_BT}\bigg)
\end{equation}
where $\mathcal{E}_\mathrm{m}$ and $\mathcal{E}_\mathrm{s}$ are the Hessian matrices of the potential energy in the (metastable) minimum and saddle point respectively. The energy difference between the minimum and saddle point is given by $\Delta U$, and $\lambda_+$ is the exponential growth rate of a small deviation from the saddle point, and $T$ is the temperature of the system.  We note that the mass $m_x$ only appears in $\lambda_+$, and the functional form of this dependence is analytically known.  

\begin{figure*}
    \centering
            \includegraphics[width=\textwidth]{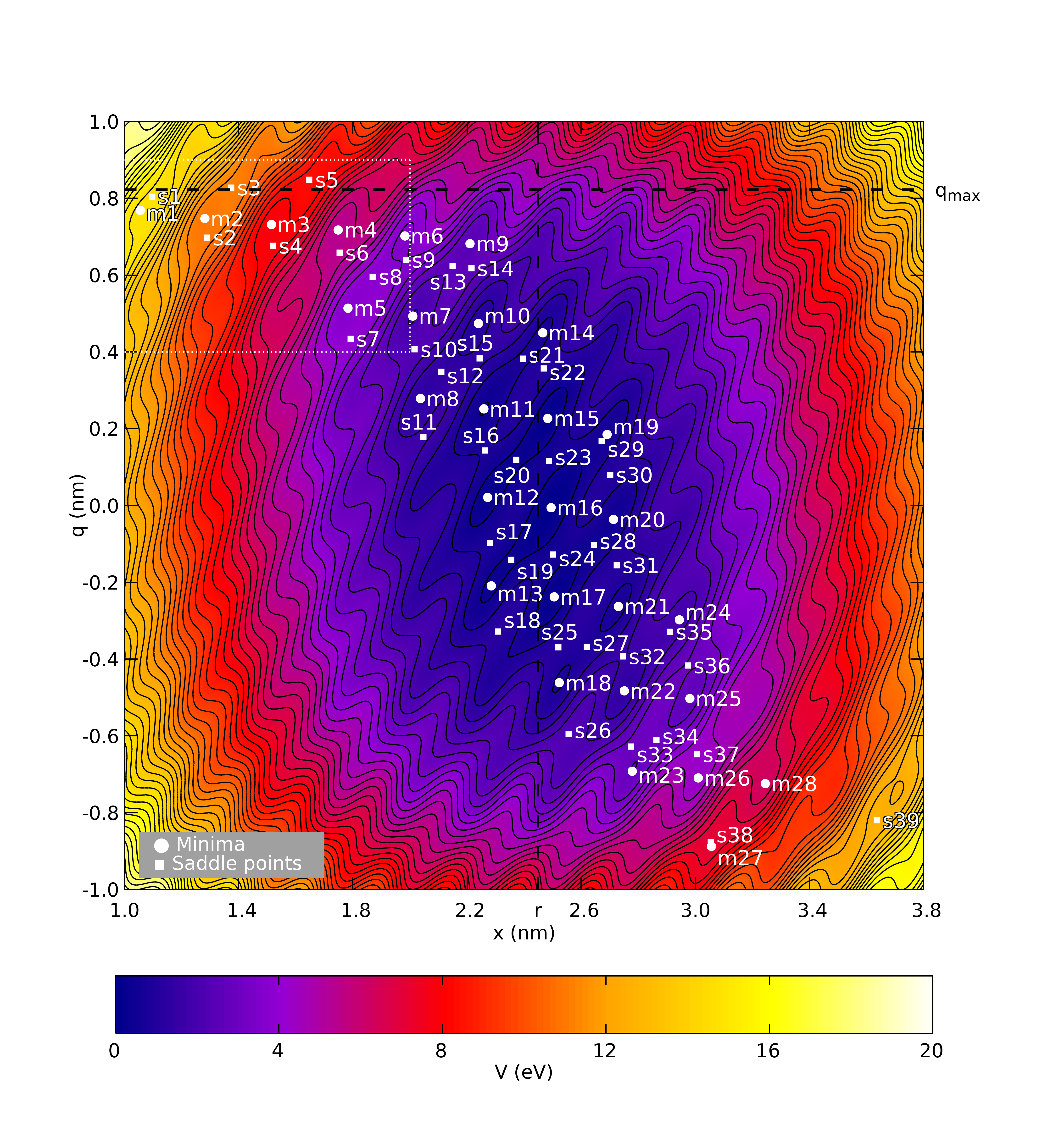}
            \caption{The potential-energy landscape for $r=2.448$~nm, with (metastable) minima and saddle points indicated. Contours are spaced every 0.4 eV to give a better guide of the shape of the landscape. The white dotted line gives context to the ranges of Figs.~\ref{fig:relaxfig}(b) and~\ref{fig:DecayPath}.
            }
            \label{fig:PotMinSad}
\end{figure*}

We numerically determine the minima and saddle points by minimising the expression $||(\partial V/\partial x,\partial V/\partial q)||^2$, and extracting any points that simultaneously meet the condition of both the partial derivatives in $x$ and $q$ are zero, $\partial V/\partial x=0$, $\partial V/\partial q=0$.  They are indicated in Fig.~\ref{fig:PotMinSad}.  We then compute the $2\times2$ Hessian matrices.
The growth of deviations from the saddle point $\lambda_+$ is obtained as the largest positive eigenvalue of the $4\times4$ dynamic matrix $\mathcal{M}$ in the saddle point, where $\mathcal{M}$ is defined by $d(x,q,\dot{x},\dot{q})/dt = \mathcal{M}\cdot (x,q,\dot{x},\dot{q})$. Applying these values to Eqn.~(\ref{eqn:kramerrate}), we then obtain the Kramers escape rates. 

What remains is the question of where the system goes after passing through a saddle point.  As can be seen from Fig.~\ref{fig:PotMinSad}, in many cases the saddle point is located almost directly between two nearby minima.  In those cases there is little question which minimum the system will decay into after passing through the saddle point.  However, there are some saddle points where this is less obvious, especially near the higher potential-energy minima.  This can be seen in Fig.~\ref{fig:PotMinSad} for example for m3 decaying via s4.  For these decays, as well as for some others, for example from m2 through s2, the reaction path is not a straight line.
In Fig.~\ref{fig:relaxfig}, this can be seen as a reversed movement in $q$.

Various minima, saddle points and their energy differences, along with the corresponding Kramers rates calculated for 300 K, are given in table~\ref{table:decaytimes}.
The most probable decay path and a sample of the decay times are indicated in Fig.~\ref{fig:MinPathLifetimes}.

\section{Comparison of decay times and paths between the analytical calculations and simulations}\label{CompCalc}

We now compare the geometric arguments and Kramers rate calculations detailed in the previous section to results from simulations as far as the available computing power allows.  It requires less statistics to determine the tendencies in the path of the tip than the relaxation time, so we first focus on this.

\subsection{Complexities in the potential-energy landscape and consequences for the decay of the tip}\label{StrangePaths}

\begin{figure*}
    \centering
            \includegraphics[width=\textwidth]{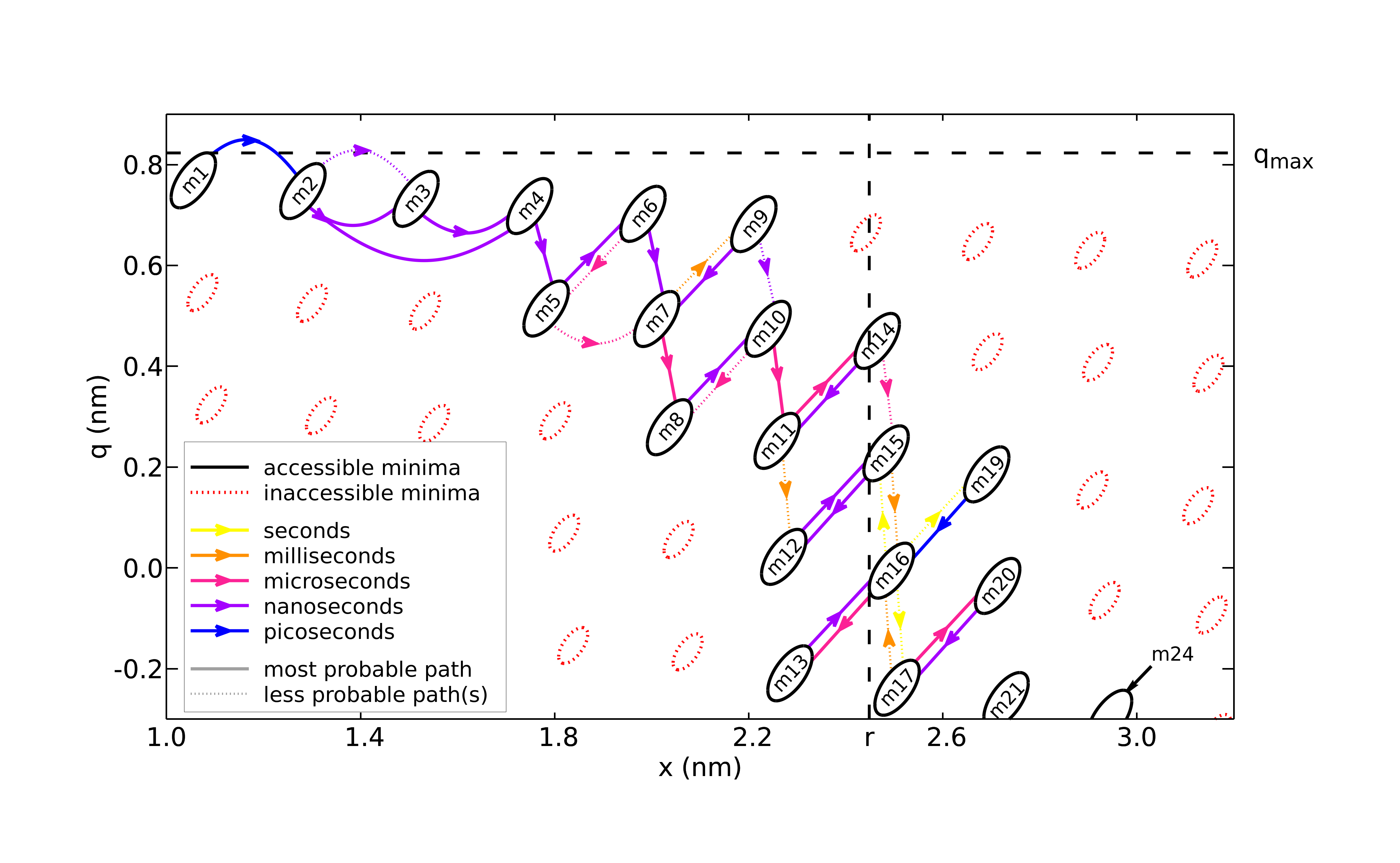}
            \caption{An illustration of the most probable paths of the tip in relation to accessible minima with metric prefixes for the decay lifetimes and probability compared to the most probable path indicated for each decay path.}
    \label{fig:MinPathLifetimes}
\end{figure*}

Examining 100 $30$ ns simulations starting from identical initial conditions described above and a fixed support at $r=2.448$~nm, we discovered several unsurprising, but nonetheless odd, paths the tip had taken. These differed from the theoretically expected paths outlined in Fig.~\ref{fig:MinPathLifetimes} and demonstrated in Fig.~\ref{fig:relaxfig}. These included 13 of which decayed from one minimum and missed several neighbouring minima before occupying another. One of these interesting paths can be seen in Fig.~\ref{fig:DecayPath}(a) where the tip moves from minimum m1 to minimum m4. Other long trajectories that bypassed minima occurred between m1 and m3, and m2 and m5. 

Another interesting decay path that occurred more frequently is a trajectory between m2 and m4 where the tip appears to occupy a minimum that does not exist, as seen in Fig.~\ref{fig:DecayPath}(b). 34 of the 100 paths examined had a trajectory similar to the one in the figure between these two minima. Even though there is no true minimum present here, the potential landscape is sufficiently shaped to allow the tip to maintain its locality for some time before finding its next minimum. This is, of course, a symptom of the geometry of the potential landscape and the support's position, as discussed in section \ref{geometry}. In comparison, 33 of the paths had similar trajectories between m2 and m3 as seen in Fig.~\ref{fig:relaxfig}(b).

Worth noting in Fig.~\ref{fig:MinPathLifetimes}, that may not be explicit in the image, is the predicted path is indicating repeated movements between some minima, indicated by two arrows. This prediction of a jiggle between two minima arrives from the calculated life times. For example, minimum m8's shortest life time, as seen in Fig.~\ref{fig:MinPathLifetimes} (and table~\ref{table:decaytimes}), is towards m10, while minimum m10's shortest life time is towards m8. This motion would repeat until the tip travels in the next most probable direction, i.e. either m11 from m8, or m11 from m10.  This behaviour is due to the structure of the landscape and should appear regardless of the precise parameters.  It is therefore a signature of the sheet dynamics that could be observed in experiments.

\subsection{Lifetimes}

\begin{figure*}
    \centering
        \begin{subfigure}[b]{0.475\linewidth}
            \includegraphics[width=\textwidth]{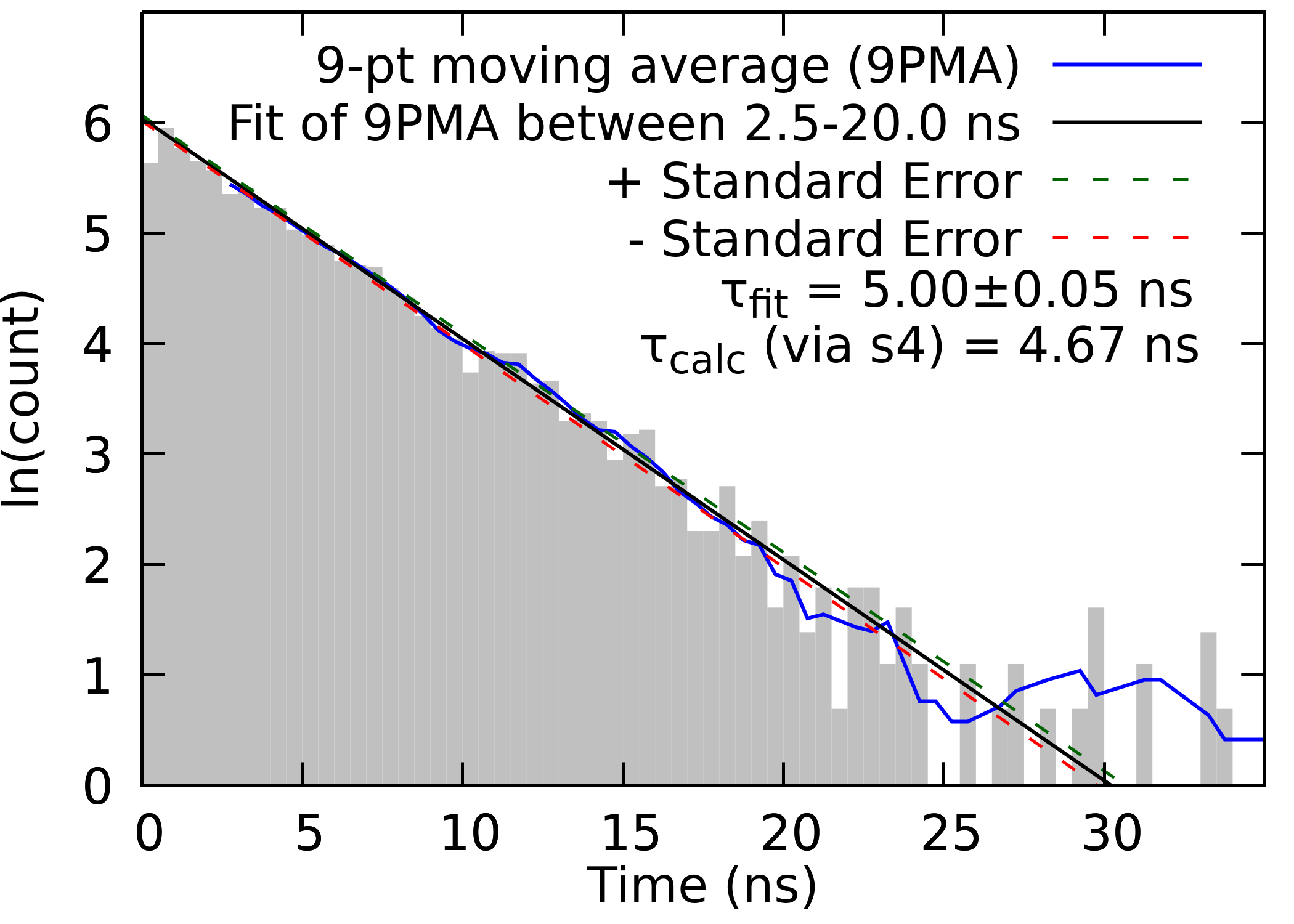}
            \caption{}
    \end{subfigure}
    \begin{subfigure}[b]{0.475\linewidth}
            \includegraphics[width=\textwidth]{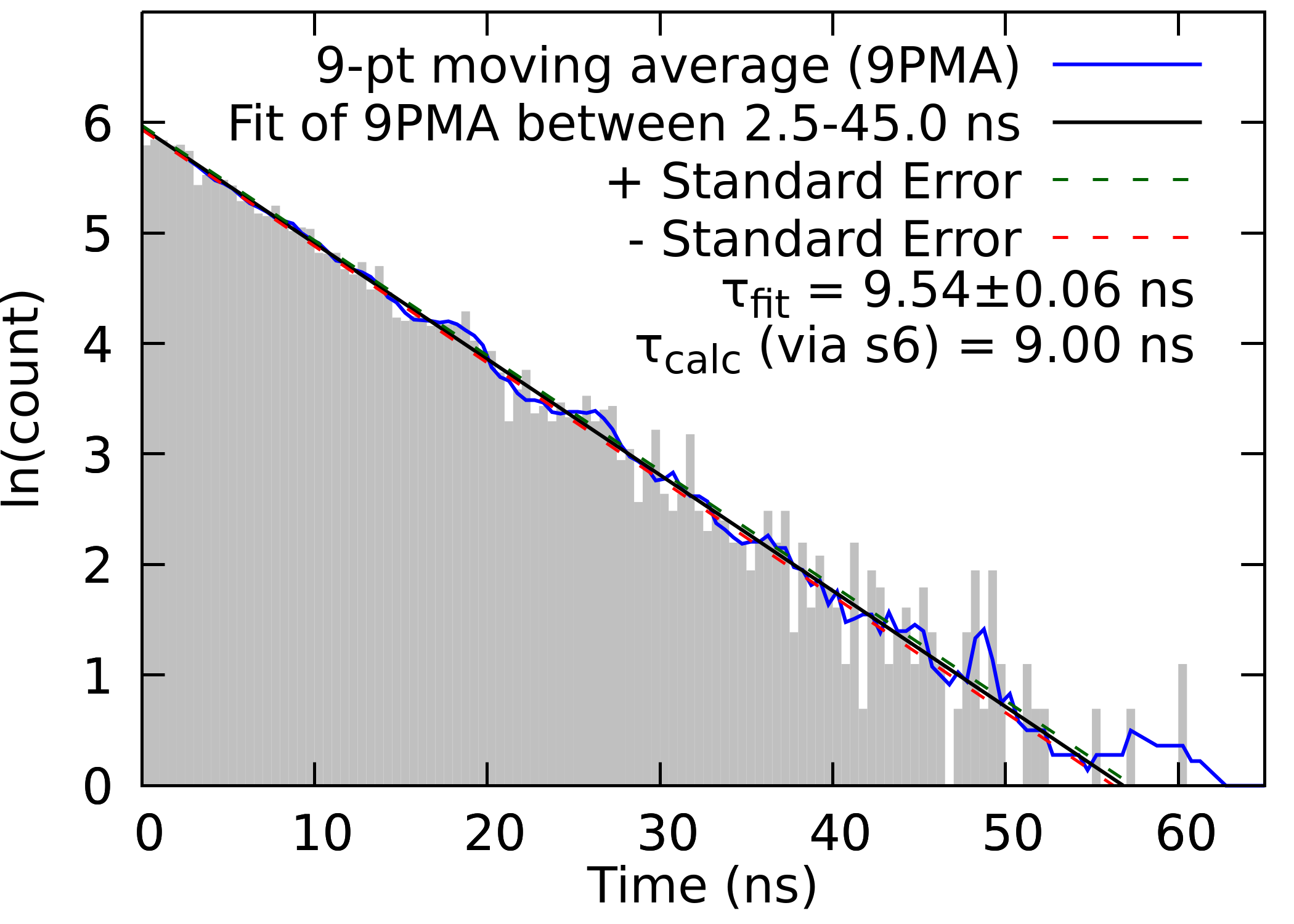}
            \caption{}
    \end{subfigure}
    \begin{subfigure}[b]{0.475\linewidth}
            \includegraphics[width=\textwidth]{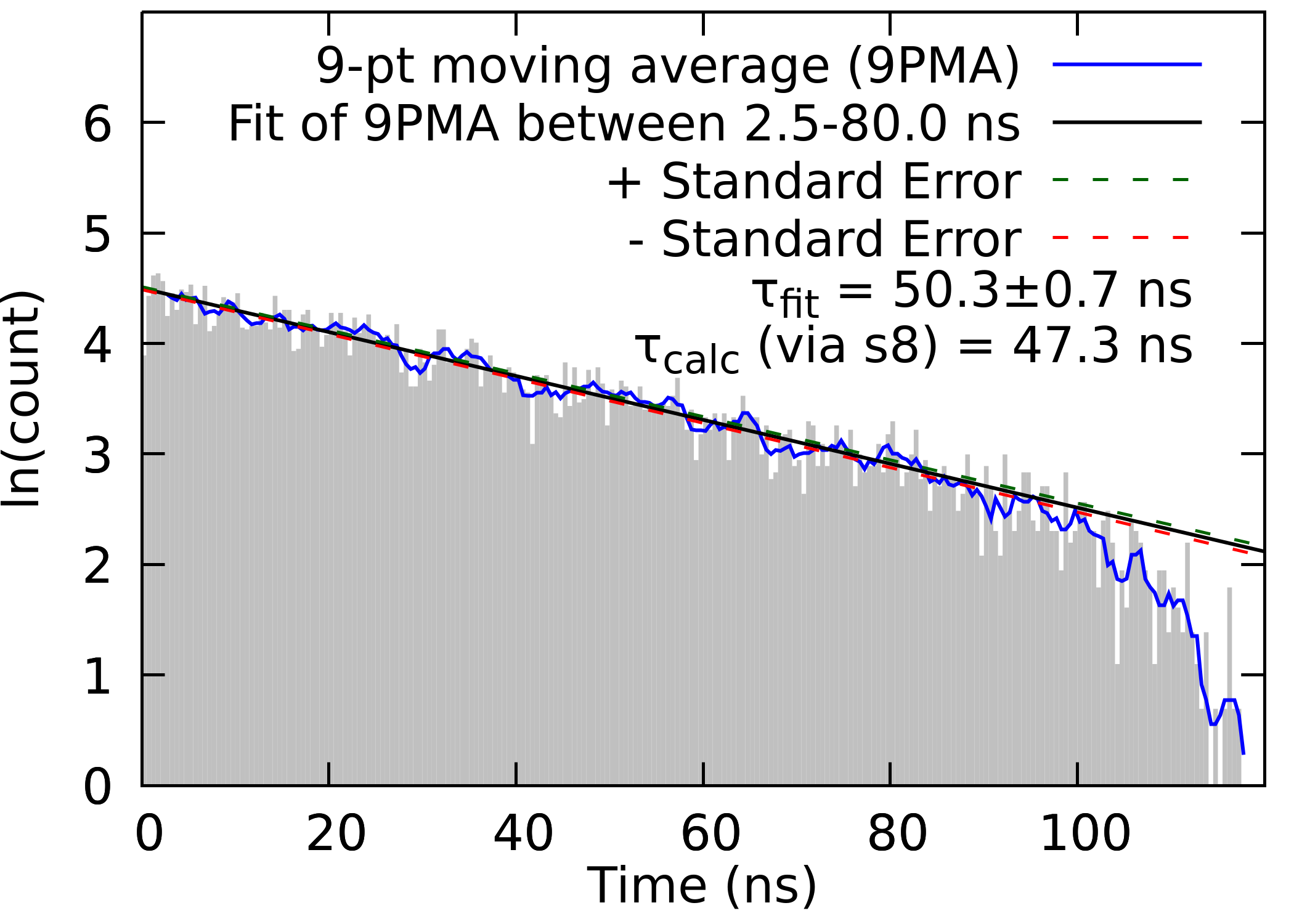}
            \caption{}
    \end{subfigure}
    \begin{subfigure}[b]{0.475\linewidth}
            \includegraphics[width=\textwidth]{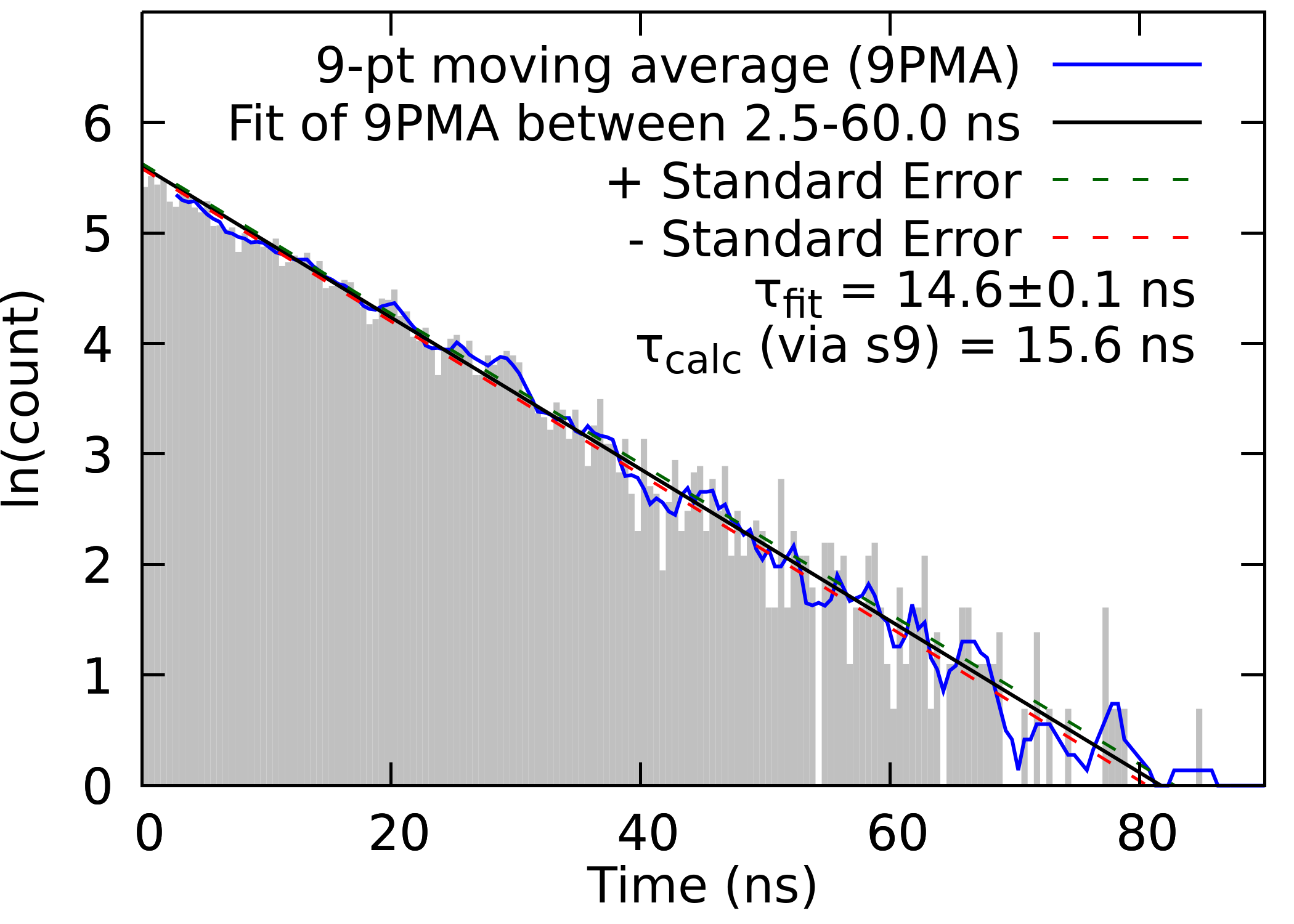}
            \caption{}
    \end{subfigure}
    \caption{Histograms of the decay times from 10 000 120 ns simulations, separated into decays between specific minima. Bin size is 0.5 ns for each histogram. The decay time from the fit and the calculated decay time via the appropriate saddle point, as in table~\ref{table:decaytimes}, are given. (a) Decay from m3 to m4. (b) Decay from m4 to m5. (c) Decay from m5 to m6. (d) Decay from m6 to m7.}
    \label{fig:Histograms}
\end{figure*}

To obtain more detailed information about the lifetimes, 10000 120~ns simulations were performed in the same manner. The decay detection algorithm described in section~\ref{sec:methods} was used to extract the time stamps and modes of decay.  In figure~\ref{fig:Histograms} we show the frequency of decay as a function of the length of time that the tip remained in a specific minimum.  The decay is exponential, as expected from any process with short-time correlations.

We fit exponential functions to a 9 point moving average (9PMA) and extract the decay rates and life times.
The results of this fitting are indicated in the figure.
The observed and calculated lifetimes are very close.  We can use the results for the Kramers decay rates for rest of the minima, that we cannot reach in the simulations, to understand what happens as the tip moves closer to the support position: the decay times increase. 
As can be seen from the more complete table of calculated life times (table~\ref{table:decaytimes}), the decay times can become quite long, in the order of s.  These long times are not achievable in the simulations, and therefore we cannot observe the full decay in simulations.

\subsection{Implications for ageing in experiments}

Signatures of this ageing and these decay rates might be visible in experiments.
While the specific decay rates and lifetimes are linked to the precise potential-energy landscape, there are general qualitative behaviours that appear regardless of the quantitative energies.

The number of decays is much larger and the path is more complex than without the presence of the sheet.  For our parameters, if the sheet is removed, there would be a maximum of 3 minima before the tip reaches the support position.  With the sheet included there are an order of magnitude more energy minima at high $q$ and positions further away from the support. This complexity gives rise to such things as the hopping back and forth that we predict will happen (for example m11 and m14 in Fig.~\ref{fig:MinPathLifetimes}). This behaviour is linked to the geometry, and does not qualitatively depend on the precise parameters. It is, therefore, likely to also appear in experiments.

Another signature of these decays is the broad range of time scales involved, from picoseconds to milliseconds (see table~\ref{table:decaytimes}).  This is a general property, and is a result of the structure of the energy landscape and the wide range of energy barriers.  This means that in experiments, where longer time scales are accessible (as opposed to in our simulations) these slower decays might be observable.  It would take the shape of a slower than exponential contact ageing.  The ageing is different from the type of contact ageing found in more complex contacts~\cite{aging2}, in that it weakens the contact and decreases the restarting friction when sliding begins again, rather than increasing it.

Although a higher number of dimensions may be truer to the complex system expected with contact ageing, it appears as though two dimensions offers enough complexity to gain insight into some of the behaviour that may appear in contact ageing.

\section{Conclusions}

We have investigated thermal effects in friction on atomically thin sheets in a simple model based on the Prandtl-Tomlinson model that reproduces the strengthening effect seen in AFM experiments.  We show that at finite temperatures there is thermolubricity, as in the normal Prandtl-Tomlinson model.  We also identify a number of thermal effects that could be observed in further experiments.

During sliding, the tip can slip to a lower distortion as well as in the sliding direction.  This can be observed as a larger than normal change in the force during slipping followed by a repeat strengthening, and a very typical shape of the force trace.

After sliding stops, the contact ages slowly due to the complex potential-energy landscape combined with thermal decay.  Due to the presence of the sheet, there are many decays with different barriers and decay rates that would not have appeared in a plain PT model.
We have performed analytical calculations to estimate the decay times that are not accessible in simulations. The initial decays are fast, but as the tip approaches the position of the support, the decay slows down.
The calculated and observed lifetimes show that there is a wide range of decay times, from several picoseconds to seconds, giving hope to the possibility that the phenomenon of relaxation can be observed in AFM experiments. In addition, there are other signatures that might be observable in experiments, such as the hopping back and forth between sets of minima on the same slanted line, which is a general feature. We also note that while there is no way to directly measure $q$ in an experiment, it is possible to obtain it indirectly by simply starting to slide again and observing the amount of strengthening that occurs.

The wide range of lifetimes and the sequence of the decays raise an interesting possibility to understand contact ageing in a general way.  While this model was designed for friction on thin sheets, it is so general that it could also be used to investigate other systems with extra degrees of freedom.
In realistic systems, many different degrees of freedom inside an asperity could play a similar role; including, but not limited to elastic deformation of the material and adsorbed layers from the atmosphere.  The model shows that even one such degree of freedom can give rise to a sequence of decay times of many different orders of magnitude, leading to a nontrivial, non-exponential ageing of the contact.

\bibliography{references}

\begin{table}[]
    \centering
    \begin{tabular}{|c|c|c|c|}\hline
        minimum & saddle point & $\Delta U/(k_\mathrm{B}T)$ & lifetime (s) \\\thickhline
        \textbf{m1} & \textbf{s1} & \textbf{0.91} & \boldmath$3.05 \times 10^{-10}$\\
        \textbf{m2} & \textbf{s2} & \textbf{4.87} & \boldmath$1.95 \times 10^{-09}$\\
        \textbf{m2} & \textbf{s3} & \textbf{9.27} & \boldmath$6.01 \times 10^{-07}$\\
        \textbf{m3} & \textbf{s4} & \textbf{5.68} & \boldmath$4.67 \times 10^{-09}$\\
        m3 & s5 & 22.8 &$4.41\times 10^{-01}$\\
        \textbf{m4} & \textbf{s6} & \textbf{6.32} & \boldmath$9.00 \times 10^{-09}$\\
        \textbf{m5} & \textbf{s7} & \textbf{13.07} & \boldmath$6.38 \times 10^{-06}$\\
        \textbf{m5} & \textbf{s8} & \textbf{6.14} & \boldmath$4.73 \times 10^{-08}$\\
        \textbf{m6} & \textbf{s8} & \textbf{13.24} & \boldmath$4.90 \times 10^{-05}$\\
        \textbf{m6} & \textbf{s9} & \textbf{6.87} & \boldmath$1.56 \times 10^{-08}$\\
        \textbf{m7} & \textbf{s10} & \textbf{13.96} & \boldmath$1.77 \times 10^{-05}$\\
        \textbf{m7} & \textbf{s13} & \textbf{19.77} & \boldmath$3.86\times 10^{-02}$\\
        m8 & s11 & 18.96 &$2.44\times 10^{-03}$\\
        \textbf{m8} & \textbf{s12} & \textbf{3.07} & \boldmath$3.16 \times 10^{-09}$\\
        \textbf{m9} & \textbf{s13} & \textbf{2.59} & \boldmath$1.57 \times 10^{-09}$\\
        \textbf{m9} & \textbf{s14} & \textbf{7.40} & \boldmath$2.70 \times 10^{-08}$\\
        \textbf{m10} & \textbf{s12} & \textbf{15.07} & \boldmath$3.83\times 10^{-04}$\\
        \textbf{m10} & \textbf{s15} & \textbf{14.55} & \boldmath$3.36 \times 10^{-05}$\\
        \textbf{m11} & \textbf{s16} & \textbf{19.76} & \boldmath$6.64\times 10^{-03}$\\
        \textbf{m11} & \textbf{s21} & \textbf{15.82} & \boldmath$9.36\times 10^{-04}$\\
        m12 & s16 & 26.76 &$8.36\times 10^{-00}$\\
        m12 & s17 & 25.34 &$1.75\times 10^{-00}$\\
        \textbf{m12} & \textbf{s20} & \textbf{6.60} & \boldmath$1.01 \times 10^{-07}$\\
        m13 & s17 & 20.59 &$1.65\times 10^{-02}$\\
        m13 & s18 & 36.00 &$4.08\times 10^{+04}$\\
        \textbf{m13} & \textbf{s19} & \textbf{2.48} & \boldmath$1.91 \times 10^{-09}$\\
        \textbf{m14} & \textbf{s21} & \textbf{2.92} & \boldmath$2.53 \times 10^{-09}$\\
        \textbf{m14} & \textbf{s22} & \textbf{15.02} & \boldmath$5.54 \times 10^{-05}$\\
        \textbf{m15} & \textbf{s20} & \textbf{8.87} & \boldmath$8.95 \times 10^{-07}$\\
        \textbf{m15} & \textbf{s23} & \textbf{20.28} & \boldmath$1.18\times 10^{-02}$\\
        \textbf{m16} & \textbf{s19} & \textbf{15.06} & \boldmath$4.83\times 10^{-04}$\\
        \textbf{m16} & \textbf{s23} & \textbf{25.85} & \boldmath$3.15\times 10^{-00}$\\
        \textbf{m16} & \textbf{s24} & \textbf{26.21} & \boldmath$4.63\times 10^{-00}$\\
        \textbf{m16} & \textbf{s29} & \textbf{22.92} & \boldmath$2.33\times 10^{-00}$\\
        \textbf{m17} & \textbf{s24} & \textbf{20.08} & \boldmath$9.44\times 10^{-03}$\\
        \textbf{m17} & \textbf{s28} & \textbf{15.13} & \boldmath$4.90\times 10^{-04}$\\
        m18 & s25 & 14.83 &$4.52 \times 10^{-05}$\\
        m18 & s27 & 7.10 &$1.34 \times 10^{-07}$\\
        \textbf{m19} & \textbf{s29} & \textbf{0.05} & \boldmath$5.87 \times 10^{-10}$\\
        m19 & s30 & 20.92 &$3.20\times 10^{-02}$\\
        \textbf{m20} & \textbf{s28} & \textbf{2.34} & \boldmath$1.76 \times 10^{-09}$\\
        m20 & s30 & 24.87 &$9.34\times 10^{-01}$\\
        m20 & s31 & 27.25 &$1.45\times 10^{+01}$\\
        m21 & s27 & 9.60 &$1.78 \times 10^{-06}$\\
        m21 & s31 & 19.49 &$4.86\times 10^{-03}$\\
        m22 & s32 & 14.33 &$2.64 \times 10^{-05}$\\
        m22 & s35 & 22.29 &$7.16\times 10^{-01}$\\
        m23 & s33 & 7.17 &$2.12 \times 10^{-08}$\\
        m23 & s34 & 6.28 &$5.05 \times 10^{-08}$\\
        m24 & s35 & 0.30 &$3.94 \times 10^{-10}$\\
        m25 & s34 & 13.35 &$5.89 \times 10^{-05}$\\
        m25 & s36 & 13.64 &$1.25 \times 10^{-05}$\\
        m26 & s37 & 6.64 &$1.25 \times 10^{-08}$\\
        m27 & s38 & 0.06 &$6.81 \times 10^{-11}$\\
        m28 & s38 & 33.05 &$4.17\times 10^{+03}$\\
        
\hline
    \end{tabular}
    \caption{Energy barriers and lifetimes for decay from various minima via specific saddle points as estimated using Kramers' escape rate theory.
    The temperature is $T=300$~K.  The lifetimes range from picoseconds to a second.  The minima, saddle points, and their labels are indicated in Fig.~\ref{fig:PotMinSad}. We do not include any decay times longer than one day. The values used in figure \ref{fig:MinPathLifetimes} are in bold in this table.}
    \label{table:decaytimes}
\end{table}

\end{document}